\documentstyle[12pt,axodraw,epsfig]{article}
\newlength{\aivwidth}   
\newlength{\tmpwidth}   

\def\lsim{\raise0.3ex\hbox{$<$\kern-0.75em\raise-1.1ex\hbox{$\sim$}}}
\def\gsim{\raise0.3ex\hbox{$>$\kern-0.75em\raise-1.1ex\hbox{$\sim$}}}
\title{\Large\bf A Simple Approximation of the One-Loop Corrected\\
Cross Section for \boldmath $e^+e^- \to W^+W^-$\unboldmath at LEP~2\footnote{
Partially supported by the Ministry of Education and Culture, Japan,
under Grant-in-Aid for basic research program C (no. 08640391),
the EC-network contract CHRX-CT94-0579 and the Bundesministerium f\"ur 
Bildung und Forschung, Bonn, Germany}}
\author{{\bf M. Kuroda}\\
Institute of Physics, Meiji Gakuin University, Yokohama 244, Japan
\bigskip\\
{\bf I. Kuss and D. Schildknecht}\\
Fakult\"at f\"ur Physik, Universit\"at Bielefeld, D-33615 Bielefeld,
Germany}
\date{hep-ph/9705294\\
BI-TP 97/15\\
May 1997}
\begin{document}
\maketitle
\vfill
\renewcommand{\abstractname}{\large\bf Abstract}
\begin{abstract}
\large
Using the $SU(2)$ gauge coupling, $g_{W^\pm} (M^2_{W^\pm})$,
at the high-energy scale of $M_{W^\pm}$, defined
by the (theoretical value of the) leptonic $W$-width, rather than using
the low-energy value, defined via the Fermi coupling, $G_\mu$, in the Born
approximation, and supplementing with Coulomb corrections and initial state
radiation, errors with respect to the exact one-loop results for the
differential cross section of $e^+e^-\to W^+W^-$ are below
1\% at LEP~2 energies at all $W^+W^-$ production angles. 
A similar procedure is suggested to incorporate leading bosonic loop
effects into four-fermion production in the fermion-loop scheme.
The resulting accuracy below 1\% is sufficient for LEP~2 experiments.
\end{abstract}
\vfill

\clearpage
\large
The process of $W$-pair production in $e^+e^-$ annihilation is presently
studied experimentally at LEP~2. In the future, it will be one of the
outstanding processes at a linear collider in the TeV energy range. 
It yields direct experimental information on the
non-Abelian couplings characteristic for the $SU(2) \times U(1)$ electroweak
theory, and it allows to put bounds on potential non-standard $Z_0W^+W^-$
and $\gamma W^+W^-$ couplings \cite{bkrs,ks}. Within the $SU(2) \times U(1)$
electroweak theory, the calculation of the radiative corrections 
to this process has received
much attention 
\cite{bohm}-\cite{been3}.\par
The exact evaluation of the full one-loop electroweak corrections leads to
complicated and lengthy expressions in terms of 
twelve $s$- and $t$-dependent form factors.
Actually, only those (three) form factors which appear in amplitudes of the
form of the Born approximation are numerically important \cite{bohm2,fleisch}. 
Unfortunately, however, no simple analytic form for these form factors,
valid at arbitrary $e^+e^-$ energies, has been given so far. For the LEP~2
energy range, however, a simple approximation has indeed been suggested
\cite{bohm2}. In this approximation, the Born approximation is evaluated
in terms of the appropriately introduced Fermi coupling, $G_\mu$, and the
high-energy electromagnetic coupling, $\alpha (s)$, and it is supplemented
by Coulomb corrections and by initial state radiation employing the structure
function method. By comparing the improved Born approximation (IBA) with
the full one-loop results, accuracies of 1.5\% to 2\% were found \cite{
been,ditt} in the angular distributions in the LEP~2 energy range.
For energies above 500 GeV, a simplification of the exact ${\cal O}
(\alpha)$ corrections has been given in terms of an asymptotic
high-energy expansion \cite{been2}.\par
It is the purpose of the present note to point out that a slight modification
of the previously suggested \cite{bohm2}
improved Born approximation for the LEP~2
energy range improves its
accuracy to values well below 1\%. Accordingly, such an approximation
will be sufficient for all practical purposes at LEP~2. Our results are
obtained by replacing the low-energy value of the $SU(2)$ gauge coupling,
$g^2_{W^\pm}(0) \equiv 4 \sqrt 2 M^2_{W^\pm} G_\mu$ previously employed
in \cite{bohm2,been,ditt}, by its high-energy value, 
defined by the leptonic $W^\pm$ width,
$g^2_{W^\pm} (M^2_{W^\pm}) \equiv 48 \pi \Gamma_l^W/M_{W^\pm}$ \cite{ditt2} 
more appropriate for
the LEP~2 energy scale\footnote{We approximate 
$g^2_{W^\pm} (s \gsim 4M^2_W)$ 
by $g^2_{W^\pm} (M_W^2)$.}. 
In essence, this approach amounts to employing a 
different renormalization scheme, defined \cite{ditt2}
by using $\Gamma_l^W$ instead of
$G_\mu$ as experimental input.
\par
The Born amplitude for the process $e^+e^- \to W^+W^-$, in the notation
of refs. \cite{been,ditt}, takes the form
\begin{equation}
{\cal M}_{Born} (\kappa, \lambda_+, \lambda_-, s,t) = 
g^2_{W^\pm} {1\over 2} \delta_{\kappa -}
{\cal M}_I + e^2 {\cal M}_Q,\label{one}
\end{equation}
where the dependence on energy and momentum transfer squared, $s$ and $t$, 
and on 
the electron and $W^\pm$ boson helicities, $\kappa = \pm 1$ and $\lambda_\pm
= 0, \pm 1$, is contained in ${\cal M}_I$ and ${\cal M}_Q$. The $SU(2)$ 
gauge coupling and the electromagnetic coupling in (\ref{one})
have been denoted by
$g_{W^\pm}$ and $e$, respectively. Even though (\ref{one}) is easily obtained 
by starting from the Feynman rules for $t$-channel neutrino and $s$-channel
$\gamma$ and $Z_0$ exchange, we prefer to gain (\ref{one}) directly from the 
electroweak theory in the $BW_3$ base, i.e. before diagonalization of
$BW_3$ mixing.\par
\begin{figure}
\epsfig{file=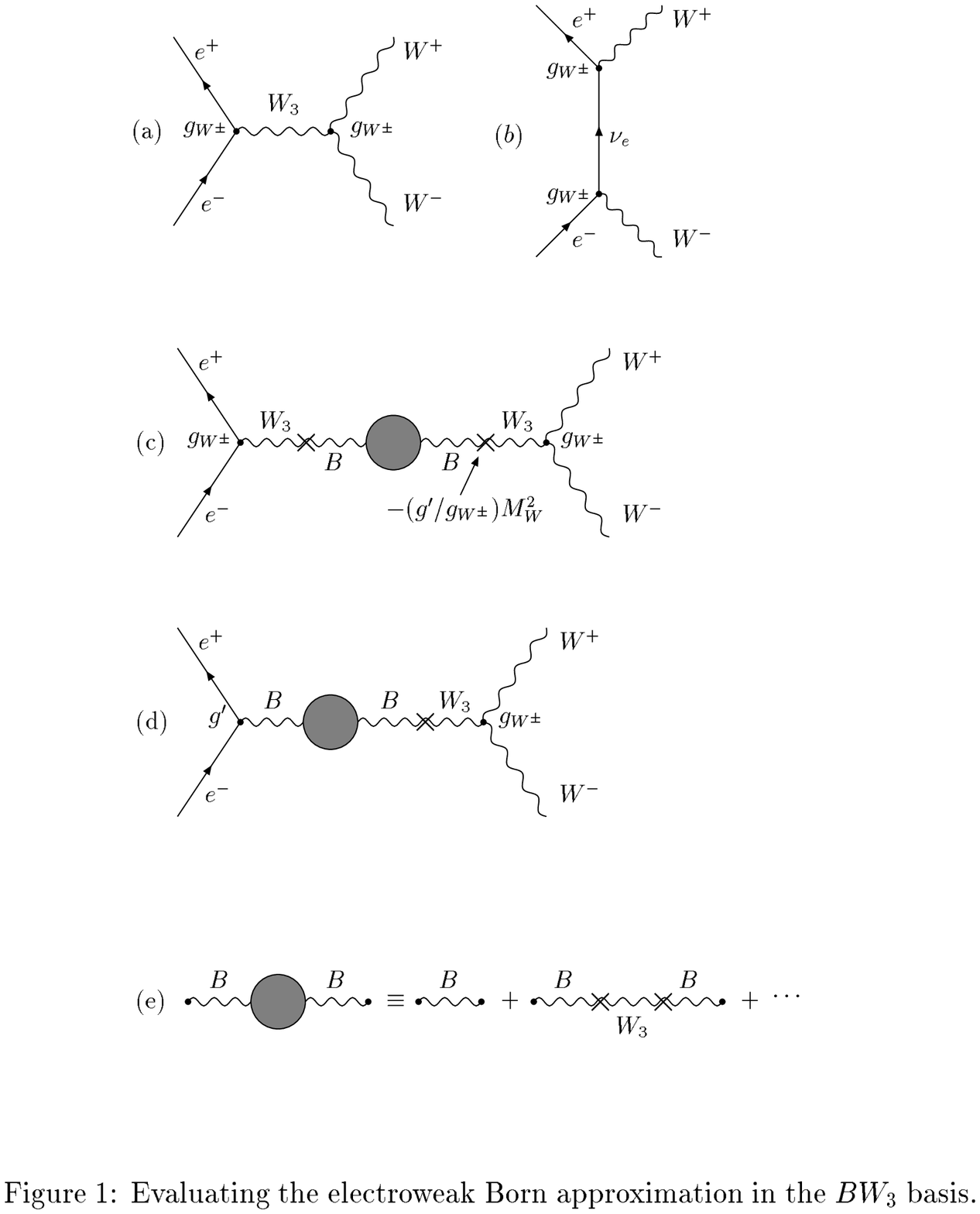}
\label{fig1}
\end{figure}
From the diagrams (a), (b) and (c) in Fig. 1, one immediately obtains
\begin{equation}
{\cal M}_I = {1 \over {s - M^2_Z}} {\cal M}_s + {1 \over t} {\cal M}_t.
\label{two}\end{equation}
One recognizes the correspondence of this expression to diagrams (a) and (b)
in Figure 1,
diagram (c) supplying the substitution $M^2_W \to M^2_Z = (1 + (g^\prime/
g_{W^\pm})^2) M^2_W$ in the $s$-channel term\footnote{In addition,
diagram (c) yields a contribution proportional to $e^2$ (via the
relation $(g')^2M_W^2=e^2M_Z^2$) which is recognized as a part of
${\cal M}_Q$.}. 
For the explicit forms of the $s$-channel and $t$-channel quantities, 
${\cal M}_s$ and ${\cal M}_t$, we refer to \cite{been}.\par
The $B$-propagator, to all orders in $BW_3$ mixing, according to diagram (e)
becomes
\begin{center}
\begin{picture}(470,40)
\Vertex(120,20){1.5}
\Photon(120,20)(150,20){2}{3}
\put(130,27){$B$}
\GCirc(163,20){13}{0.5}
\Photon(176,20)(206,20){2}{3}
\Vertex(206,20){1.5}
\put(186,27){$B$}
\put(216,17){=}
\put(235,15){\large $\frac{\displaystyle s-M_W^2}{\displaystyle s(s-M_Z^2)}$,}
\put(451,15){\large (3)}
\end{picture}
\end{center}
\addtocounter{equation}{1}
where $M^2_B \equiv (g^\prime/g_{W^\pm})^2 M^2_W$ for the square of the
$B$-boson mass and $-(g^\prime /
g_{W^\pm})$ $\cdot M^2_W$ for the mixing strength were used. 
For right-handed electrons, only the
$B$-coupling (to the weak hypercharge current) of diagram (d) is relevant,
implying, with (3) and $g^{\prime 2} M^2_W = e^2M^2_Z$, that 
\begin{equation}
{\cal M}_Q = -{M_Z^2 \over s(s-M_Z^2)} {\cal M}_s.
\label{four}\end{equation}
This expression holds equally well for left-handed electrons, where 
contributions from the diagrams (d) and (a)+(c) make up
one half of ${\cal M}_Q$ each.

We feel that the above derivation of (\ref{one}) illuminates in the most
straight-forward manner the decomposition of (\ref{one}) into a weak 
$SU(2)$ and an
electromagnetic piece, where the latter one for right-handed electrons is
entirely induced by the $B$-boson coupling to the hypercharge current. 
Moreover, the origin of the double-pole structure in (\ref{four}) as a result of
$BW_3$ mixing becomes immediately obvious. The double pole leads to 
a suppression of the amplitude (\ref{four}) relative to (\ref{two}) 
at high energies, 
which in the case of longitudinal $W^\pm$
bosons is compensated, however, by the longitudinal polarization vectors.

The calculation of the cross section for $e^+e^- \to W^+W^-$ from (\ref{one})
requires the specification of a scale at which the $SU(2)$ gauge coupling
$g^2_{W^\pm}$ and the electromagnetic coupling $e^2$ are to be
defined. As $W$ pairs are
produced at LEP~2 at energies of $2M_{W^\pm} \lsim \sqrt s \sim 200 GeV$, it
is natural to choose a high-energy scale, such as $\sqrt s$. We expect that
it is sufficiently accurate to use the scale $M_W \simeq M_Z$ instead of
$\sqrt s$. Accordingly, we choose \cite{burk}
\begin{equation}
\left({{e^2} \over {4 \pi}}\right)^{-1} = 
\alpha^{-1} (M^2_Z) = 128.89 \pm 0.09,
\label{five}\end{equation}
for the electromagnetic coupling, and define $g^2_{W^\pm} (M^2_W)$ by the
leptonic width of the $W^\pm$ \cite{ditt2}\footnote{We note that
$g^2_{W^\pm}(M_{W^\pm}^2)$ as defined by (\ref{six}) differs from
$g^2_{W^\pm}(M_{W^\pm}^2)$ as defined by (6) in ref. \cite{ditt2}
by the factor $r\equiv 1+c_0^2\cdot 3\alpha/4\pi$, where
$c_0^2\cdot 3\alpha/4\pi\simeq 1.34\cdot 10^{-3}$. The factor $r$ 
corresponds to
the factor $1+3\alpha/4\pi$ in the $Z^0$ width, where it is
conventionally introduced in order to explicitly separate photon
radiation from all other electroweak one-loop corrections. The
introduction of the factor $r$ in the $W^\pm$ width allowed \cite{ditt2}
to correctly
define the magnitude of isospin breaking by one-loop weak interactions
when passing from the charged boson coupling $g_{W^\pm}(M_{W^\pm}^2)$ to
the neutral boson coupling $g_{W^0}(M_Z^2)$, while keeping the usual
convention of separating photon radiation from other loop corrections in
the expression for the $Z^0$ width. For the purposes of the present
paper we have removed the factor $r$. This amounts to including all
one-loop radiative corrections in $g_{W^\pm}^2(M_{W^\pm}^2)$ as defined
in (\ref{six}). All qualitative conclusions of the present work remain
the same if $\Delta y^{SC}\simeq 3.3\cdot 10^{-3}$ from (\ref{14}),
corresponding to
(\ref{six}), (\ref{seven}) and (\ref{eight}), is replaced by $\Delta
y^{SC}\simeq 4.6\cdot 10^{-3}$ from ref. \cite{ditt2}.}
\begin{equation}
g^2_{W^\pm} (M^2_{W^\pm}) = 48 \pi {{\Gamma_l^W}\over {M_{W^\pm}}}.
\label{six}\end{equation}
Expressing $\Gamma_l^W$ in terms of the Fermi coupling,
\begin{equation}
\Gamma^W_l = {{G_\mu M^3_W} \over {6 \sqrt 2 \pi (1 + 
{\Delta y}^{SC})}},
\label{seven}\end{equation}
including one-loop corrections, one finds that $g^2_{W^\pm} 
(M_W^2)$ differs from
the gauge coupling, $g^2_{W^\pm} (0)$, defined from $\mu$-decay, by the
correction factor $(1 + {\Delta y}^{SC})^{-1}$,
\begin{equation}
g^2_W (M^2_{W^\pm}) = {{g^2_{W^\pm} (0)} \over 
{1 + {\Delta y}^{SC}}}.
\label{eight}\end{equation}
The ``scale-change (SC)'' part, ${\Delta y}^{SC}$, of the coupling
parameter $\Delta y$ of ref. \cite{ditt2}, takes care of the change 
in scale between
$\mu$-decay and $W^\pm$-decay. It is given by
\begin{equation}
{\Delta y}^{SC} = \Delta y^{SC}_{ferm} + {\Delta y}^{SC}_{bos},
\label{nine}\end{equation}
where the fermionic part, $\Delta y^{SC}_{ferm}$,
is essentially due to contributions arising from
light fermion loops in the $W^\pm$ propagator. For $m_t \to \infty$, it is
given by
\begin{equation}
\Delta y^{SC}_{ferm} \vert_{m_t \to \infty} = - {{3 \alpha (M^2_Z)} \over
{4 \pi s^2_0}} \simeq - 8.01 \cdot 10^{-3},
\label{ten}\end{equation}
while for $m_t = 180$ GeV,
\begin{equation}
\Delta y^{SC}_{ferm} \vert_{m_t = 180 GeV} = - 7.79 \times 10^{-3}.
\label{11}\end{equation}
This negative contribution to ${\Delta y}^{SC}$
is largely compensated by the bosonic one, ${\Delta y}^{SC}_{bos}$, which
is practically independent\footnote{The parameter 
${\Delta y}^{SC}_{bos} = 11.13 \times 10^{-3}$ for $M_H = 100$ GeV, 
while
${\Delta y}^{SC}_{bos} = 11.08 \times 10^{-3}$ for $M_H = 300$ GeV and
$\Delta y^{SC}_{bos} = 11.07 \times 10^{-3}$ for $M_H = 1000$ GeV 
\cite{ditt2}.}
of the Higgs boson mass and is given by\footnote{Note that $\Delta
y^{SC}_{bos}$(this paper)  $=\Delta y^{SC}_{bos}$(Table 1 in ref. 
\cite{ditt2}) $-1.34\cdot 10^{-3}$. Compare footnote 3.}
\begin{equation}
{\Delta y}^{SC}_{bos} = 11.1 \times 10^{-3}.
\label{12}\end{equation}
Accordingly, the SU(2) coupling, $g^2_{W^\pm} (M_{W^\pm}^2)$, to be used when
evaluating the cross section for the process $e^+e^- \to W^+W^-$
in the Born approximation (\ref{one}) is given by
\begin{equation}
g^2_{W^\pm} (M^2_{W^\pm}) = {{4 \sqrt 2 G_\mu M^2_{W^\pm}} \over {1 + 
{\Delta y}^{SC}}},
\label{13}\end{equation}
with a correction term, due to scale change, whose magnitude,
\begin{equation}
{\Delta y}^{SC} = 3.3 \times 10^{-3},
\label{14}\end{equation}
is practically independent of the precise values of $m_t$ and $M_H$.
Even though there is such a strong cancellation between fermions and bosons
in ${\Delta y}^{SC}$, thus implying the fairly small value of
${\Delta y}^{SC}$ in (\ref{14}), the correction induced by
${\Delta y}^{SC}$ will be seen
to be decisive for providing the announced accuracy, better than 1\%, in the
expression for the $W^\pm$ pair-production cross section.\par
Including the Coulomb correction and the initial state radiation (ISR)
in soft photon approximation, the
improved Born approximation for the differential cross section takes the form
\begin{eqnarray}
\left({{d \sigma} \over {d \Omega}}\right)_{IBA} &=& {\beta \over {64 \pi^2 s}}
\left\vert {{2 \sqrt 2 G_\mu M^2_W} \over {1 + {\Delta y}^{SC}}}
{\cal M}^\kappa_I \delta_{\kappa^-} + 4 \pi\alpha (M^2_Z) {\cal M}^\kappa_Q 
\right\vert^2\cr
&&+\left({{d\sigma} \over {d \Omega}}\right)_{Coul} (1 - \beta^2)^2 + 
\left({{d\sigma} \over
{d\Omega}}\right)_{ISR}.\label{15}
\end{eqnarray}
The only difference of the present work with respect to refs. 
\cite{bohm2,been,ditt}
consists in the inclusion of ${\Delta y}^{SC}$ which introduces
according to (\ref{13})
the appropriate high-energy scale for the $SU(2)$-coupling strength.
We note the dependence of $(d\sigma/d\Omega)_{ISR}$ in (\ref{15})
on the choice of the
photon splitting scale, $Q^2$, inherently connected with the structure
function method. This method consists of 
evaluating the leading logarithmic QED corrections
thus all contributions proportional to
$(\alpha/\pi)\ln(m_e^2/Q^2)$.

The deviation of the differential cross section in the
improved Born approximation (without the 
${\Delta y}^{SC}$ correction)
from the full one-loop result normalized by the Born cross section,
$\Delta_{IBA}$, was worked out numerically in refs. \cite{bohm2,been,ditt}. 
The introduction
of ${\Delta y}^{SC}$ in (\ref{15}) simply amounts to an additive correction
to $\Delta_{IBA}$. This additive correction, $\delta \Delta_{IBA}$, 
is calculated by evaluating
\begin{equation}
\delta \Delta_{IBA} = {{({{d\sigma} \over {d\Omega}})_{IBA~({\Delta y}^
{SC} \not= 0)} - ({{d\sigma} \over {d\Omega}})_{IBA~({\Delta y}^
{SC} = 0)}}\over {({{d\sigma} \over {d\Omega}})_{Born}}},\label{16}
\end{equation}
and the quality of the approximation (\ref{15}) to the full one
loop result is accordingly
quantified by
\begin{equation}
\Delta_{IBA}+\delta\Delta_{IBA}.\label{17}
\end{equation}
We note that the magnitude of $\delta \Delta_{IBA}$ may easily be estimated
due to the fact that the ${\cal M}_I$ part dominates the cross section 
(\ref{15}).
Indeed, neglecting ${\cal M}_Q$ in (\ref{15}), one obtains from
(\ref{16}),
\begin{equation}
\delta \Delta_{IBA} \simeq - 2 {\Delta y}^{SC} = -0.66\% \label{18}
\end{equation}
as a rough estimate. This value will be somewhat enhanced or diminished,
depending on whether the interference term of the ${\cal M}_I$ with the
${\cal M}_Q$ term in (\ref{15}) is negative 
(as in the forward region) or positive
(as in the backward region). 

The results for $\delta\Delta_{IBA}$ are given in Table \ref{table1} and
the percentage quality of our improved Born approximation (\ref{15}),
$\Delta_{IBA}+\delta\Delta_{IBA}$, is compared with $\Delta_{IBA}$. The
values for $\Delta_{IBA}$ are taken from Table 4 in ref. \cite{been}.
They are based on the choice of $Q^2\equiv s$ for the photon splitting
scale $Q^2$. One observes that indeed the deviation of the unpolarized
cross section from the full
one-loop results is improved to less than 1\% as a consequence of
introducing ${\Delta y}^{SC}$ in (\ref{15}). 
As the scale $Q^2$ is by no means theoretically uniquely fixed, we also show,
in Table \ref{table2}, the results for the different choice of 
$Q^2=M_W^2$.
Even though the uncorrected quality of the approximation,
$\Delta_{IBA}$\footnote{We thank S. Dittmaier for providing us with
the values
of $\Delta_{IBA}$ for the photon splitting scale $Q^2=M_W^2$.}, is better 
in this case than for $Q^2=s$, the inclusion
of $\delta\Delta_{IBA}$ 
again leads to an improvement of the quality of the approximation also
in this case, and values below approximately 0.5\% are reached.
In other words, the qualitative improvement in the approximation
(\ref{15}), obtained by introducing $\Delta y^{SC}\neq 0$, is
independent of the choice of $Q^2$.

For completeness,
in Tables \ref{table1} and \ref{table2}, 
we also present the results for the cross section for left-handed
electrons, which obviously do not differ much from the results for the
unpolarized cross section, since the right-handed cross section is suppressed
by about two orders of magnitude compared with the left-handed one. The
right-handed cross section by itself is obviously unaffected by introducing
${\Delta y}^{SC}$.\par
A final comment concerns the inclusion of the decay of the $W^\pm$'s into
fermion pairs which has to be incorporated into a completely realistic
description of the process of $W^\pm$ pair production. A gauge-invariant
description of the process $e^+e^-\to 4$ fermions at one-loop order was
recently given in the fermion-loop approximation \cite{loopscheme}.
In this connection it seems worth while to come back to the
decomposition of $\Delta y^{SC}$ into fermion-loop and bosonic
contributions in (\ref{nine}), (\ref{11}) and (\ref{12}).
We note that taking into account fermion-loop contributions only, the
estimate (\ref{18}) becomes 
\begin{equation}
\delta\Delta_{IBA}\vert_{ferm}\simeq -2\Delta y^{SC}_{ferm} 
\vert_{m_t = 180 GeV} \simeq +1.56\%, 
\label{19}\end{equation}
and the total deviation $\Delta_{IBA}+\delta\Delta_{IBA}$ of (\ref{15})
from
the full one-loop results (using $\Delta_{IBA}\simeq 1.2\%$ from Table
\ref{table1}) rises to values of the
order of 2.8\% for the total cross section.
Therefore, we expect that four-fermion production
evaluated in the fermion-loop approximation \cite{loopscheme}
is also enhanced by as much as approximately 2.8\% relative to the (so
far unknown) outcome of a full one-loop calculation incorporating
bosonic loops as well. It is gratifying that a simple procedure of
taking into account bosonic loops to improve the results of the
fermion-loop calculations of four-fermion production immediately
suggests itself. We suggest to approximate bosonic loop corrections by
carrying out the substitution
\begin{equation}
G_\mu\to G_\mu/(1+\Delta y^{SC}_{bos})
\label{20}\end{equation}
with $\Delta y^{SC}_{bos}=11.1\cdot 10^{-3}$ in the four-fermion
production amplitudes evaluated in the fermion-loop approximation.
Substitution (\ref{20}) practically amounts to using
$g_{W^\pm}(M_{W^\pm}^2)$ in four-fermion production as well.
With substitution (\ref{20}) it is indeed to be expected that the 
deviation of four-fermion production in the fermion-loop
scheme will be diminished from the above estimated value of
$\simeq 2.8\%$ to a value below 1\%.

In summary, the simple procedure of introducing the $SU(2)$ gauge coupling
$g_{W^\pm} (M^2_W)$ 
at the high-energy scale, approximated by $s \simeq M^2_W$,
or in other words, by introducing a renormalization scheme, in which the 
$SU(2)$ coupling is defined by the (theoretical value of the radiatively
corrected) leptonic width
of the $W$-boson, allows one to incorporate most of the electroweak radiative
corrections to the process of $e^+e^- \to W^+W^-$ in the LEP~2 energy range
of $\sqrt s\, \lsim\, 200 GeV$.
Adding the Coulomb corrections and the initial state radiation in the
leading logarithmic approximation provides
a scheme which approximates the full one-loop results with an accuracy
better than 1\%. 
Moreover, we suggest a simple recipe to approximately incorporate
bosonic corrections into four-fermion production calculations, which so
far are available in the fermion-loop approximation only.
The overall accuracy thus obtained should be sufficient for the analysis 
of $W$ pair production at LEP~2.

\section*{Acknowledgement}
The authors would like to thank S. Dittmaier and H. Spiesberger for
useful discussions.

\clearpage
\begin{table}
\begin{center}
\vspace{-5mm}
\begin{tabular}{|r|r|r|c||r|r|c|}
\hline
\multicolumn{1}{|c|}{angle}&\multicolumn{3}{c||}{unpolarized}&
\multicolumn{3}{c|}{left-handed}\\
\hline
&\multicolumn{1}{c|}{$\Delta_{IBA}$}&$\delta\Delta_{IBA}$&
\multicolumn{1}{c|}{$\Delta_{
IBA}+\delta\Delta_{IBA}$}&
$\Delta_{IBA}$&$\delta\Delta_{IBA}$&
\multicolumn{1}{c|}{$\Delta_{IBA}+\delta\Delta_{IBA}$}\\
\hline\hline
\multicolumn{7}{|c|}{$\sqrt{s}=161$ GeV}\\
\hline
\multicolumn{1}{|c|}{total}&1.45&-0.72&0.73&1.45&-0.72&0.73\\
10&1.63&-0.73&0.90&1.63&-0.73&0.90\\
90&1.44&-0.72&0.72&1.44&-0.72&0.72\\
170&1.26&-0.70&0.56&1.26&-0.70&0.56\\
\hline\hline
\multicolumn{7}{|c|}{$\sqrt{s}=165$ GeV}\\
\hline
\multicolumn{1}{|c|}{total}&1.27&-0.71&0.56&1.28&-0.71&0.57\\
10&1.67&-0.74&0.93&1.67&-0.74&0.93\\
90&1.17&-0.71&0.46&1.18&-0.71&0.47\\
170&0.75&-0.67&0.08&0.77&-0.67&0.10\\
\hline\hline
\multicolumn{7}{|c|}{$\sqrt{s}=175$ GeV}\\
\hline
\multicolumn{1}{|c|}{total}&1.26&-0.71&0.55&1.28&-0.71&0.57\\
10&1.71&-0.75&0.96&1.71&-0.75&0.96\\
90&1.03&-0.69&0.34&1.06&-0.70&0.36\\
170&0.59&-0.62&-0.03&0.69&-0.63&0.06\\
\hline\hline
\multicolumn{7}{|c|}{$\sqrt{s}=184$ GeV}\\
\hline
\multicolumn{1}{|c|}{total}&1.02&-0.70&0.32&1.06&-0.71&0.35\\
10&1.57&-0.75&0.82&1.57&-0.75&0.82\\
90&0.67&-0.68&-0.01&0.72&-0.69&0.03\\
170&0.10&-0.58&-0.48&0.32&-0.64&-0.32\\
\hline\hline
\multicolumn{7}{|c|}{$\sqrt{s}=190$ GeV}\\
\hline
\multicolumn{1}{|c|}{total}&1.24&-0.70&0.54&1.28&-0.71&0.57\\
10&1.67&-0.74&0.93&1.67&-0.75&0.92\\
90&0.95&-0.68&0.27&1.01&-0.69&0.32\\
170&0.58&-0.57&0.01&0.83&-0.59&0.24\\
\hline\hline
\multicolumn{7}{|c|}{$\sqrt{s}=205$ GeV}\\
\hline
\multicolumn{1}{|c|}{total}&1.60&-0.70&0.90&1.65&-0.71&0.94\\
10&1.77&-0.74&1.03&1.77&-0.74&1.03\\
90&1.55&-0.66&0.89&1.64&-0.68&0.96\\
170&1.61&-0.53&1.08&1.94&-0.56&1.38\\
\hline
\end{tabular}
\end{center}
\caption{
The Table shows the quality of the improved Born approximation (IBA) for
the total (defined by integrating over $10^0\lsim\vartheta\lsim 170^0$)
and the differential cross section (for $W^-$-production
angles $\vartheta$ of $10^0,90^0$ and $170^0$) for $e^+e^-\to W^+W^-$ 
at various
energies for unpolarized and left-handed electrons. The quantity
$\Delta_{IBA}$ denotes the percentage deviation of the IBA from the full
one-loop result, the numerical results being taken from ref. \cite{been}. Our
correction, $\delta\Delta_{IBA}$, as well as the final accuracy,
$\Delta_{IBA}+\delta\Delta_{IBA}$, of our IBA
are given in the second and third column, respectively.
The photon splitting scale entering the cross section for initial state
bremsstrahlung is chosen as $Q^2=s$.}
\label{table1}
\end{table}

\clearpage
\begin{table}
\begin{center}
\begin{tabular}{|r|r|r|c||r|r|c|}
\hline
\multicolumn{1}{|c|}{angle}&\multicolumn{3}{c||}{unpolarized}&
\multicolumn{3}{c|}{left-handed}\\
\hline
&\multicolumn{1}{c|}{$\Delta_{IBA}$}&$\delta\Delta_{IBA}$&
\multicolumn{1}{c|}{$\Delta_{
IBA}+\delta\Delta_{IBA}$}&
$\Delta_{IBA}$&$\delta\Delta_{IBA}$&
\multicolumn{1}{c|}{$\Delta_{IBA}+\delta\Delta_{IBA}$}\\
\hline\hline
\multicolumn{7}{|c|}{$\sqrt{s}=161$ GeV}\\
\hline
\multicolumn{1}{|c|}{total}&0.97&-0.72&0.25&0.97&-0.72&0.25\\
10&1.14&-0.73&0.41&1.14&-0.73&0.41\\
90&0.95&-0.72&0.23&0.96&-0.72&0.24\\
170&0.78&-0.70&0.08&0.78&-0.70&0.08\\
\hline\hline
\multicolumn{7}{|c|}{$\sqrt{s}=165$ GeV}\\
\hline
\multicolumn{1}{|c|}{total}&0.77&-0.71&0.06&0.78&-0.71&0.07\\
10&1.17&-0.74&0.43&1.17&-0.74&0.44\\
90&0.67&-0.71&-0.04&0.68&-0.71&-0.03\\
170&0.25&-0.67&-0.42&0.27&-0.67&-0.40\\
\hline\hline
\multicolumn{7}{|c|}{$\sqrt{s}=175$ GeV}\\
\hline
\multicolumn{1}{|c|}{total}&0.70&-0.71&-0.01&0.73&-0.71&0.02\\
10&1.17&-0.75&0.42&1.17&-0.75&0.42\\
90&0.48&-0.69&-0.21&0.51&-0.70&-0.19\\
170&0.05&-0.62&-0.57&0.15&-0.63&-0.48\\
\hline\hline
\multicolumn{7}{|c|}{$\sqrt{s}=184$ GeV}\\
\hline
\multicolumn{1}{|c|}{total}&0.43&-0.70&-0.27&0.47&-0.71&-0.24\\
10&0.99&-0.75&0.24&0.99&-0.75&0.24\\
90&0.09&-0.68&-0.59&0.14&-0.69&-0.55\\
170&-0.48&-0.58&-1.06&-0.26&-0.64&-0.90\\
\hline\hline
\multicolumn{7}{|c|}{$\sqrt{s}=190$ GeV}\\
\hline
\multicolumn{1}{|c|}{total}&0.63&-0.70&-0.07&0.67&-0.71&-0.04\\
10&1.07&-0.74&0.33&1.07&-0.75&0.32\\
90&0.35&-0.68&-0.33&0.41&-0.69&-0.28\\
170&-0.02&-0.57&-0.59&0.23&-0.59&-0.36\\
\hline\hline
\multicolumn{7}{|c|}{$\sqrt{s}=205$ GeV}\\
\hline
\multicolumn{1}{|c|}{total}&0.94&-0.70&0.24&0.99&-0.71&0.28\\
10&1.11&-0.74&0.37&1.12&-0.74&0.38\\
90&0.90&-0.66&0.24&0.99&-0.68&0.31\\
170&0.96&-0.53&0.43&1.28&-0.56&0.72\\
\hline
\end{tabular}
\end{center}
\caption{
As Table \ref{table1}, but for the photon splitting scale
$Q^2=M_W^2$.}
\label{table2}
\end{table}

\end{document}